\documentclass{aa}

\usepackage{graphicx}

\begin{document}

\title{Inferences from pulsational amplitudes and phases for multimode $\delta$ Sct star FG Vir}

\author{J. Daszy\'nska-Daszkiewicz$^{1,2}$, W. A. Dziembowski$^{2,3}$,
A. A. Pamyatnykh$^{2,4,5}$,  \\M. Breger$^5$, W. Zima$^5$, G.
Houdek$^6$}

\offprints{J. Daszy\'nska-Daszkiewicz
\email{daszynska@astro.uni.wroc.pl}}

\institute{{1} Astronomical Institute, Wroc{\l}aw University,
ul. Kopernika 11, 51-622 Wroc{\l}aw, Poland\\
 {2} Copernicus Astronomical Center, ul.
Bartycka 18, 00-716 Warsaw, Poland\\
 {3} Warsaw University
Observatory, Al. Ujazdowskie 4, Warsaw, Poland\\
 {4} Institute of
Astronomy, Russian Academy of Science, Pyatnitskaya 48, 109017
Moscow, Russia\\
 {5} Institute of Astronomy, University of Vienna,
T\"urkenschanzstr. 17, 1180 Vienna, Austria\\
 {6} Institute of Astronomy, University of Cambridge, Cambridge CB3 0HA, UK\\
    }

\date{Received ...; accepted ...}

\abstract{ We combine photometric and spectroscopic data on twelve
modes excited in FG Vir to determine their spherical harmonic
degrees, $\ell$, and to obtain constraints on the star model. The
effective temperature consistent with the mean colours and the pulsation
data is about 7200K. In six cases, the $\ell$ identification
is unique with above 80\,\% probability. Two modes are identified
as radial. Simultaneously with $\ell$, we determine a complex
parameter $f$ which probes subphotospheric stellar layers.
Comparing its values with those derived from models assuming
different treatment of convection, we find evidence that
convective transport in the envelope of this star is inefficient.
\\
}
   \titlerunning{Inferences from pulsational amplitudes and phases for FG Vir}
   \authorrunning{Daszy\'nska-Daszkiewicz et al.}

\maketitle

\section{Introduction}

FG Vir is the most studied $\delta$ Sct star. After the Sun it is
the Main Sequence star with the largest number of eigenfrequencies
measured. In spite of efforts (e.g.  Breger et al. 1995, Guzik \&
Bradley 1995, Viskum et al. 1998, Breger et al. 1999, Templeton et
al. 2001 ), we still do not have a good seismic model of this
object. Not much has been learnt so far from this rich frequency
data. The main obstacle is the lack of revealing features in the
oscillation spectrum. For a few dominant peaks identification of
the spherical harmonic degree, $\ell$, have been suggested.
However, even with this few $\ell$ values the task of
disentangling the spectrum appears formidable. It is frustrating
that so far the progress in amplitude resolution resulting in ever
growing number of detected modes does not help. It seems that the
science will be served better if we focus on information contained
in a few peaks for which we have reliable information on
amplitudes and phases of the light variation in various photometric
passbands as well as of the radial velocity variation.
The problem with low-amplitude peaks is that, rather than just
representing missing component of low-degree multiplets, they may
also correspond to moderate-degree modes of unknown azimuthal order.

The identification of $\ell$ values is of the highest priority as
a first step towards a unique mode identification. The most widely
used tool for $\ell$ determination has been the amplitude ratio
vs. phase differences diagrams in two passbands.
\begin{figure*}
 \includegraphics[width=175mm,height=13cm]{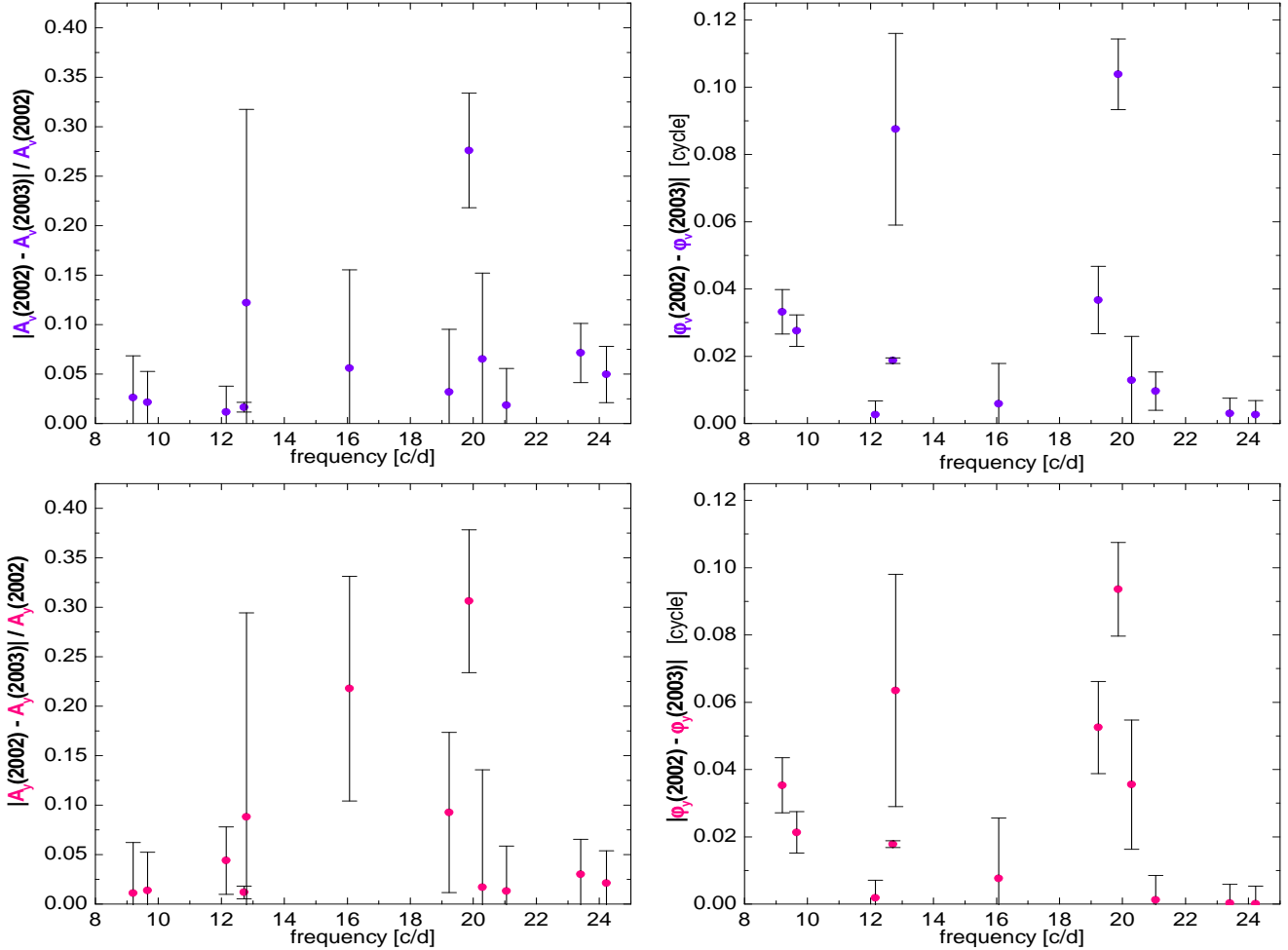}
  \caption{Changes in the observed photometric amplitudes (left panels)
   and phases (right panels) of FG Vir light curves between
   the 2002 and 2003 seasons. The upper panels are for the $v$ Str\"omgren
   passband and the lower panels are for the $y$ passband.}
  \label{fig:diff}
\end{figure*}

The usefulness of these diagrams is limited because mode
localization depends not only on $\ell$ but also on a certain
complex parameter describing the relative perturbation of the
bolometric flux. The parameter, which we denote $f$, may be
determined by solving the linear nonadiabatic oscillation problem
for the stellar model. Unfortunately, the calculated $f$ values
are unreliable because they depend on stellar parameters and, more
importantly, on the treatment of convection. In this paper, we
follow an alternative approach (Daszy\'nska-Daszkiewicz,
Dziembowski, Pamyatnykh, 2003 (Paper I), 2004), which does not require
any assumptions about convection because $f$ is determined
together with  $\ell$ from observations. The inferred value of $f$
is of interest in itself.

The determination of $\ell$ and $f$ values from pulsation
amplitudes and phases requires atmospheric models calculated for
the star mean photospheric parameters. Nonetheless, we  still call
such $f$'s {\it empirical}. There is an uncertainty in these
parameters, which must be taken into account, but it is far less
severe than in the {\it theoretical} $f$'s derived as solutions of
the linear nonadiabatic oscillation problem.

A comparison of the empirical and theoretical $f$'s yields a new
seismic probe. The value of $f$ is determined in the layers, where
the thermal time scale is of the order of pulsation period. These
are subphotospheric layers and they are poorly probed by
frequencies. Thus, $f$'s must be regarded as a supplementary probe
of star interiors.

Our paper is constructed as follows. In Sect.\,2. we review recent
observational data on FG Vir. Identification of the spherical
harmonic degrees, $\ell$, as well as the $f$ values for twelve
dominant modes in the oscillation spectrum is presented in Sect.\,3.
Constraints on models of convection based on the $f$ values are
derived in Sect.\,4.

\section{Observations}

Three recent, extensive photometric campaigns on FG~Vir were
undertaken in the years 2002, 2003 and 2004 by the Delta Scuti
Network (Breger et al. 2003, 2005), while spectroscopic
observations were obtained in 2002. For the analysis undertaken in
this paper, we have only used the photometric data from 2002,
rather than adopting the combined 2002--2004 solution, which has
lower observational uncertainties. The reason is presented in
Fig.\,1, which demonstrates that small year-to-year changes in the
amplitudes and phases may exist. Such changes might be intrinsic
to the star due to amplitude variability or observational due to
missing frequencies of low amplitude.  
In spite of some annual amplitude and phase variability,
the amplitude ratios, $A_v/A_y$, and phase shifts, $\varphi_v-\varphi_y$,
show no annual variations beyond the statistical uncertainties
expected from the calculated photometric errors.
Thus, the amplitude and phase variability does
not affect the mode identification.
In Fig.\,2 we present amplitude ratio {\it vs.} phase difference
diagrams, which have been traditionally used for mode degree
identification. For certain modes, we see significant differences
in the positions determined from the 2002 and 2003 data.
Indeed, if there are amplitude and/or phase changes, we may obtain
an incorrect result for $\ell$ from data which are not simultaneous.
It appears safer to rely only on the  photometric and spectroscopic
data obtained during 2002.

\begin{figure*}
 \includegraphics[width=175mm]{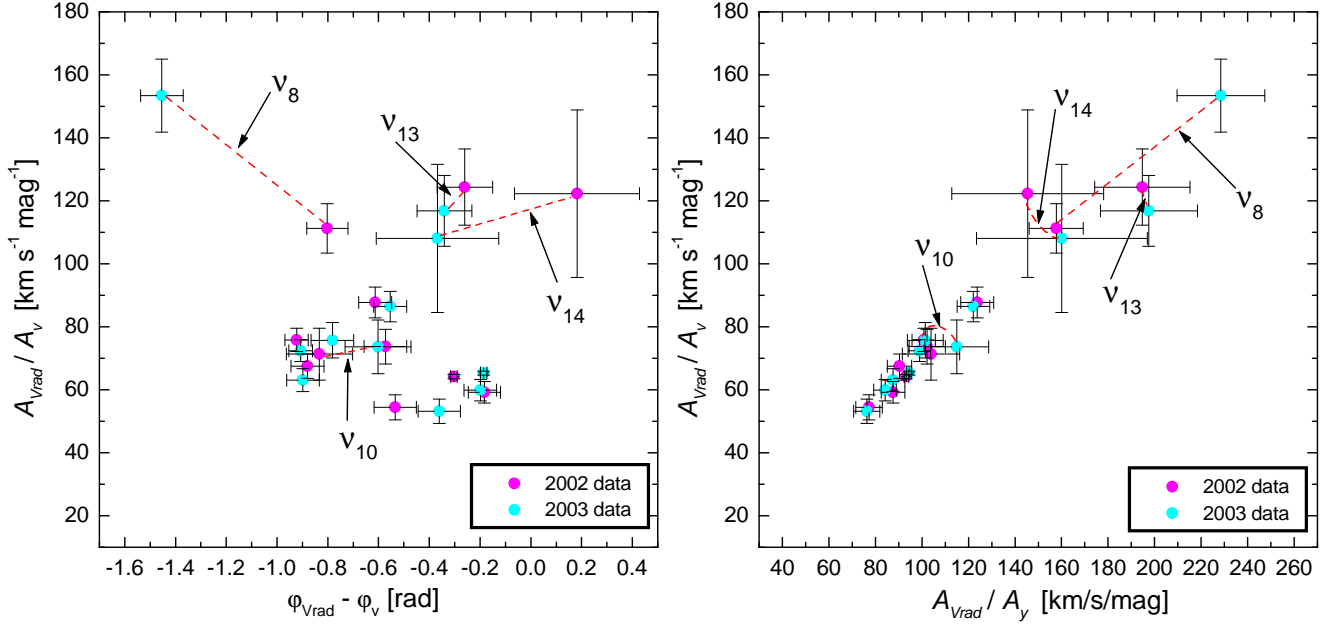}
  \caption{  The amplitude ratio vs. phase difference diagram (left panel)
             and the amplitude ratio vs. amplitude ratio diagram (right panel)
             for the dominant modes in FG Vir oscillation spectrum.
             The diagrams are based on the radial velocity data taken in 2002
           and the two seprate sets of photometric data taken in 2002 and 2003.}
  \label{fig:diag}
\end{figure*}

\section{Inferring $\ell$ and $f$ from observations}

\subsection{The method}

The method is described in detail in Paper I. Here we give only a brief outline.

The complex photometric amplitudes for a number of passbands,
$\lambda$, are written in the form of the following set of linear
observational equations
\begin{equation}
{\cal D}_{\ell}^{\lambda} ({\tilde\varepsilon} f)
+{\cal E}_{\ell}^{\lambda} {\tilde\varepsilon} = A^{\lambda},
\end{equation}
where
\begin{equation}
{\tilde\varepsilon} = \varepsilon Y^m_{\ell}(i,0),
\end{equation}
\begin{equation}
{\cal D}_{\ell}^{\lambda} = \frac14 b_{\ell}^{\lambda}
\frac{\partial \log ( {\cal F}_\lambda |b_{\ell}^{\lambda}| ) }
{\partial\log T_{\rm{eff}}}
\end{equation}
\begin{equation}
{\cal E}_{\ell}^{\lambda} =  b_{\ell}^{\lambda}
\left[ (2+\ell )(1-\ell ) - \left( \frac{\omega^2 R^3}{G M} + 2 \right)
\frac{\partial \log ( {\cal F}_\lambda
|b_{\ell}^{\lambda}| ) }{\partial\log g} \right]
\end{equation}
Having spectroscopic observations we can supplement the above set
with the expression for the radial velocity (the first moment,
${\cal M}_1^{\lambda}$)
\begin{equation}
{\rm i}\omega R \left( u_{\ell}^{\lambda}
+ \frac{GM}{R^3\omega^2} v_{\ell}^{\lambda} \right)
{\tilde\varepsilon}={\cal M}_1^{\lambda}
\end{equation}
Symbols in Eqs. (1)$-$(5) have the following meaning.
$\varepsilon$ is a complex parameter fixing mode amplitude and
phase, $i$ is the inclination angle, and
$b_{\ell}^{\lambda},~u_{\ell}^{\lambda},~v_{\ell}^{\lambda}$ are
limb-darkening-weighted  disc averaging factors. The quantity
${\cal F}_\lambda(T_{\rm eff},g)$ denotes the monochromatic flux,
which is determined from a static atmosphere model. The model
enters through ${\cal D}_{\ell}^{\lambda}$, ${\cal
E}_{\ell}^{\lambda}$ and through the disc averaging factors, which
contain the limb-darkening coefficients. The atmosphere model
depends on $T_{\rm eff}$, $\log g$, metallicity parameter, [m/H],
and on the microturbulence velocity, $\xi_{\rm t}$. In principle, the
latter quantity is subject to pulsational variations but we will
ignore perturbation of $\xi_{\rm t}$ in this paper.

Each passband, $\lambda$, yields r.h.s. of equations (1).
Measurements of radial velocity yield r.h.s. of equation (5). With
data from two photometric passbands  and from spectroscopy, we
have three complex linear equations for two complex unknowns:
${\tilde\varepsilon}$ and $({\tilde\varepsilon}f)$. Note that
$|\varepsilon f|$ is the relative amplitude of the luminosity
variations. The equations are solved by the LS method for
specified $\ell$ values. The $\ell$ determination is based on
$\chi^2(\ell)$ minima.

\subsection{Mean stellar parameters}

In our preliminary study of FG Vir (Daszy\'nska-Daszkiewicz et al. 2004),
we adopted the following ranges for the stellar parameters:
$M=1.8-1.9 M_{\odot}$, $\log T_{\rm eff}={\bf 3.866-3.884}$ and $\log
L/L_{\odot}=1.12-1.22$. These ranges are consistent with the mean
Str\"omgren photometric data, the Hipparcos parallax and the
evolutionary tracks for the Population~I composition. As the
standard, we used atmospheric models of Kurucz (1998) but we also
considered models from different sources. A good fit of the
pulsational amplitudes and phases was obtained at $\log T_{\rm
eff}=3.875$ but unfortunately, it was artificial. It resulted from
a non-smooth behaviour of the flux derivatives with respect to
$T_{\rm eff}$. Smooth derivatives were determined with the use of
much denser tabular data from NEMO.2003 models (Nendwich et al.
2004), but the fit at $\log T_{\rm eff}=3.875$ was very bad. A
satisfactory fit was reached at much lower temperature, $\log
T_{\rm eff}\approx3.82$, which lead to mean colours inconsistent
with observations.

We began the present investigation with searching for stellar parameters
that are consistent with the mean colours and that lead to a satisfactory fit
of the pulsational amplitudes and phases for the dominant peak in the
oscillation spectrum ($\nu_1=12.716$ c/d). We relied on the NEMO.2003
stellar atmosphere models. It turned out that the fit for consistent
stellar parameters may be possible by adjusting the metallicity
parameter, [m/H], and the microturbulence velocity, $\xi_{\rm t}$. The
goal was achieved either by increasing [m/H] from 0.0 to +0.2 or
$\xi_{\rm t}$ from 2 to 4 km$\,$s$^{-1}$. How an increase of $\xi_{\rm t}$
leads to the agreement in the effective temperature is illustrated in Fig.\,3.
The values of $\chi^2$, plotted in this figure and quoted later in
this paper, are calculated per degree of freedom, which is 2 in
our case. Of the two options, we chose the increasing $\xi_{\rm t}$
because there is spectroscopic evidence for a solar metal
abundance in FG Vir (Mittermayer \& Weiss 2003). Moreover, the
same authors suggest the higher value of $\xi_{\rm t}$.
\begin{figure}
 \includegraphics[width=84mm]{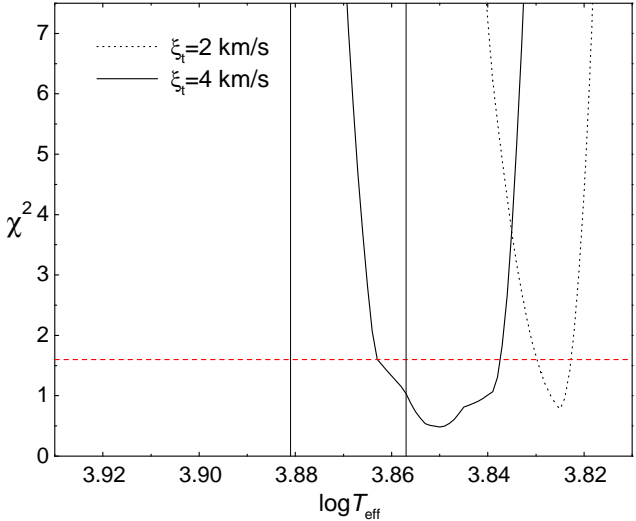}
  \caption{ Behaviour of $\chi^2 (T_{\rm eff})$ from the fits of pulsation amplitudes
   and phases for the dominant peak, $\nu_1$, which was identified as $\ell=1$. Atmospheric models
   were calculated with [m/H]=0.0 and two indicated values of the
   microturbulence velocity, $\xi_{\rm t}$.
   The horizontal line at $\chi^2=1.6$ corresponds to 80\,\% confidence level.
   The vertical lines marks the range of $T_{\rm eff}$ from mean colours.}
 \label{fig:lident}
\end{figure}

The basic stellar parameters adopted for the present investigation
were derived from the mean photometric indices and the Hipparcos
parallax using NEMO.2003 models. We arrived at the following
values, $\log T_{\rm eff}=3.869\pm 0.012$, $\log
L/L_{\odot}=1.170\pm 0.055$. For the evaluation of the effective
gravity we derived masses from our evolutionary tracks.

We should stress that also a satisfactory agreement between mean
colours and pulsational data may be achieved at similar [m/H] and
$\xi_{\rm t}$ with Kurucz's models calculated without overshooting in
the atmosphere.

The radiative flux derivatives needed for our method were determined by
numerical differentiation of the tabular data from NEMO.2003
models. The limb-darkening coefficients were taken from Barban et
al. (2003).
\begin{figure*}
 \includegraphics[width=175mm, height=15cm]{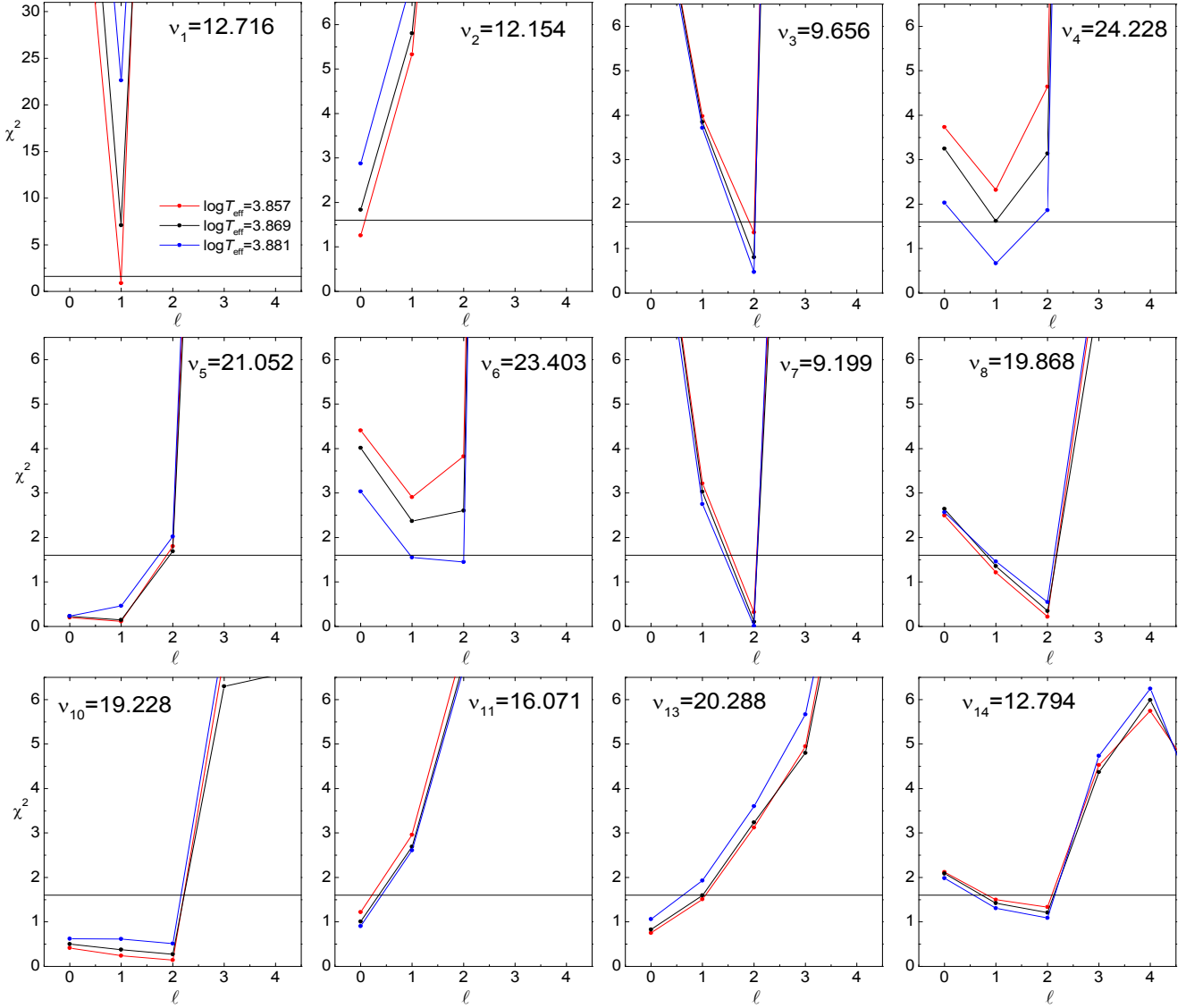}
  \caption{The values of $\chi^2$ as a function of $\ell$ for twelve dominant peaks
           in the oscillation spectrum. The assumed parameters for the atmosphere
            models are: $\log g=4.0$, [m/H]=0.0, $\xi_{\rm t}=4$ km$\,$s$^{-1}$, the three values
            of $\log T_{\rm eff}$ are given in the legend. The horizontal line at $\chi^2=1.6$
            corresponds to the 80\,\% confidence level.  }
 \label{fig:lident}
\end{figure*}

\begin{table*}\def~{\hphantom{0}}
  \begin{center}
  \centerline{Table 1. Possible identification of $\ell$ (with 80\,\% probability)
   within the accepted $T_{\rm eff}$ range}
  \label{tab_ident}
  \begin{tabular}{lccc}\hline
   $\nu$ [c/d] & our identification & Viskum et al. & Breger et al.\\
               & from phot.\&spec.  &     (1998)     &     (1999)   \\
      \hline
   $\nu_1$= 12.716    &   $\ell=1$   &  $\ell=1$ & $\ell=1$   \\
   $\nu_2$= 12.154    &   $\ell=0$    & $\ell=0$ & $\ell=0$   \\
   $\nu_3$=  9.656    &   $\ell=2$   & $\ell=2$ & $\ell=1,2$ \\
   $\nu_4$= 24.228    &   $\ell=1$    & $\ell=1$ & $\ell=1,2$ \\
   $\nu_5$= 21.052    &  $\ell=1,0$   & $\ell=2$ & $\ell=2$   \\
   $\nu_6$= 23.403    &  $\ell=2,1$   & $\ell=0$ & $\ell=0,1$ \\
   $\nu_7$=  9.199    &  $\ell=2$     & $\ell=2$ & $\ell=2$   \\
   $\nu_8$= 19.868    &  $\ell=2,1$   & $\ell=2$ & $\ell=2$   \\
  $\nu_{10}$= 19.228  & $\ell=2,1,0$ & $-$  & $-$\\
  $\nu_{11}$= 16.071  &  $\ell=0$    &  $-$ & $-$ \\
  $\nu_{13}$= 20.288  &  $\ell=0,1$  &  $-$ & $-$ \\
  $\nu_{14}$= 12.794  &  $\ell=2,1$  &  $-$ & $-$ \\
\hline
  \end{tabular}
 \end{center}
\end{table*}

\subsection{Identification of spherical harmonic degrees}

We applied the method described in Sect.\,3.1 to the twelve
dominant modes in FG Vir pulsation. For all of them, we have both
photometric and spectroscopic data. We use only observations made
in 2002 because radial velocity measurements are only from 2002.
In the present application the radial velocity data are essential
because we have data only for two photometric passbands and three
is the minimum if we want to rely on the pure photometric version
of our method.

In Fig.\,4 we plot $\chi^2$ as a function of $\ell$. With the
adopted 80\, \% confidence level, corresponding to $\chi^2=1.6$, a
unique $\ell$ identification is not always possible. The $\chi^2$
minima  for the dominant peak and the majority of the remaining
ones favour the lowest values of $\log T_{\rm eff}$ in the allowed
range. There are two exceptions, the $\nu_4$ and $\nu_6$ peaks,
which prefer the highest $\log T_{\rm eff}$. Of course there is
only one effective temperature and we believe that it is close to
$\log T_{\rm eff}\approx 3.86$, because the $\nu_1$ is the
dominant mode and its amplitudes and phases are most accurate.

In Table 1 we compare our new $\ell$ identification with earlier
attempts. The agreement is very good. We accepted only the $\ell$ values
leading to $\chi^2\le 1.6$, which results in a 80\,\% probability of the
correct identification. With this criterion we could assign
unique $\ell$ values to the six frequencies.

It is significant that $\ell\ge 3$ is excluded in all twelve cases
at a safe confidence level. We know very little about nonlinear
development of multimode instability. If only the disc averaging
effect were responsible for mode selection then there would be a
fair chance for detecting modes with $\ell\ge 3$. It would be so
because, in the range of observed frequencies, the amplitude
reduction between $\ell=1$ and $\ell=4$, for instance, is only
about six, whereas all modes from $\nu_2$  to $\nu_{14}$  have
amplitudes by factor 5 to 20 lower than the $\nu_1$ mode.

The identification of the $\nu_2$ and $\nu_{11}$ peaks as radial modes
looks firm and promising. The probability that one of the modes has
$\ell>0$ is less than 5 \%. The frequency ratio of 0.756 is
not far from the expected ratio between the fundamental mode and
the first overtone. A closer look, however, reveals that achieving
a close match is not easy. The corresponding model values range
from 0.773 to 0.779 depending which opacity data are used in the
stellar model. The lower value is obtained with the OPAL (Iglesias
\& Rogers 1996) and the higher one with the OP data (Seaton \&
Badnell 2004, Badnell et al. 2005, Seaton 2005). It seems very
difficult to reconcile the ratio with models ignoring effects of
rotation. As Daszy\'nska-Daszkiewicz et al. (2002) have shown,
rotation couples the even $\ell$ modes and it is possible that for
instance $\ell=4$ mode may mimic a radial mode. It is beyond the
scope of the present paper to examine this possibility, but we
plan to do it in the future.

\begin{figure*}
 \includegraphics[width=175mm, height=8cm]{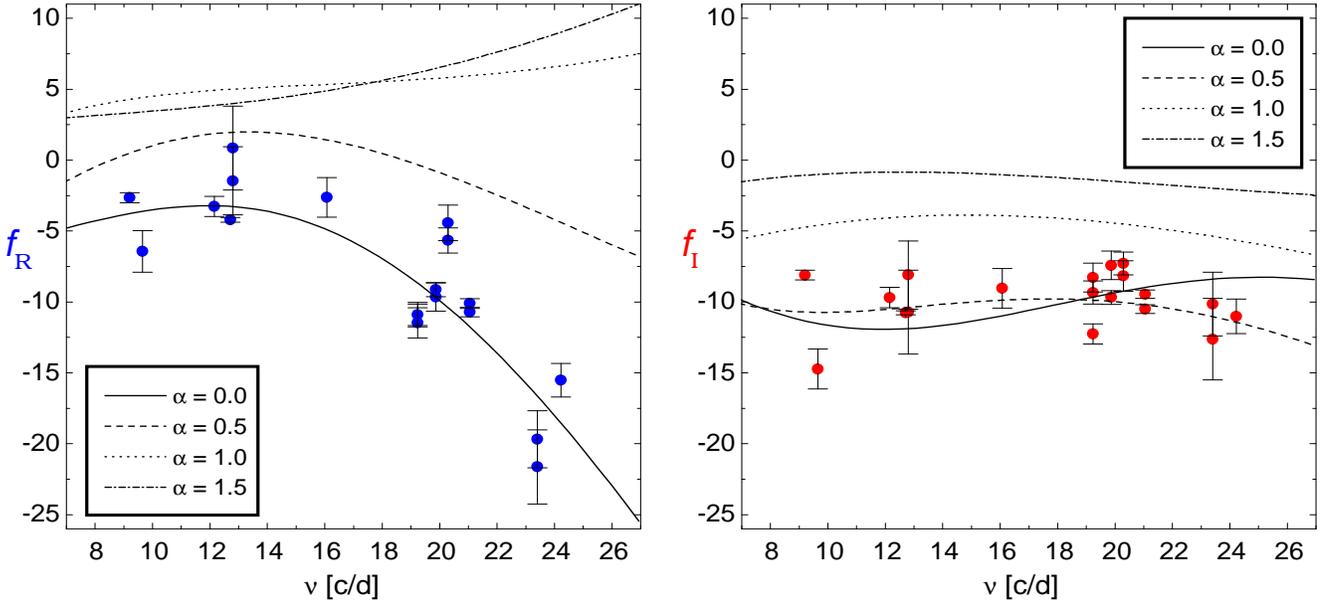}
  \caption{Comparison of the empirical values of  $f$
        (dots with error bars) with the theoretical ones calculated
        for four values of the MLT parameter, $\alpha$, adopting convective
        flux freezing approximation. The real and imaginary part of
        $f$ are shown in the left and the right panels, respectively.}
  \label{fig:fparam}
\end{figure*}
\begin{figure*}
 \includegraphics[width=175mm, height=8cm]{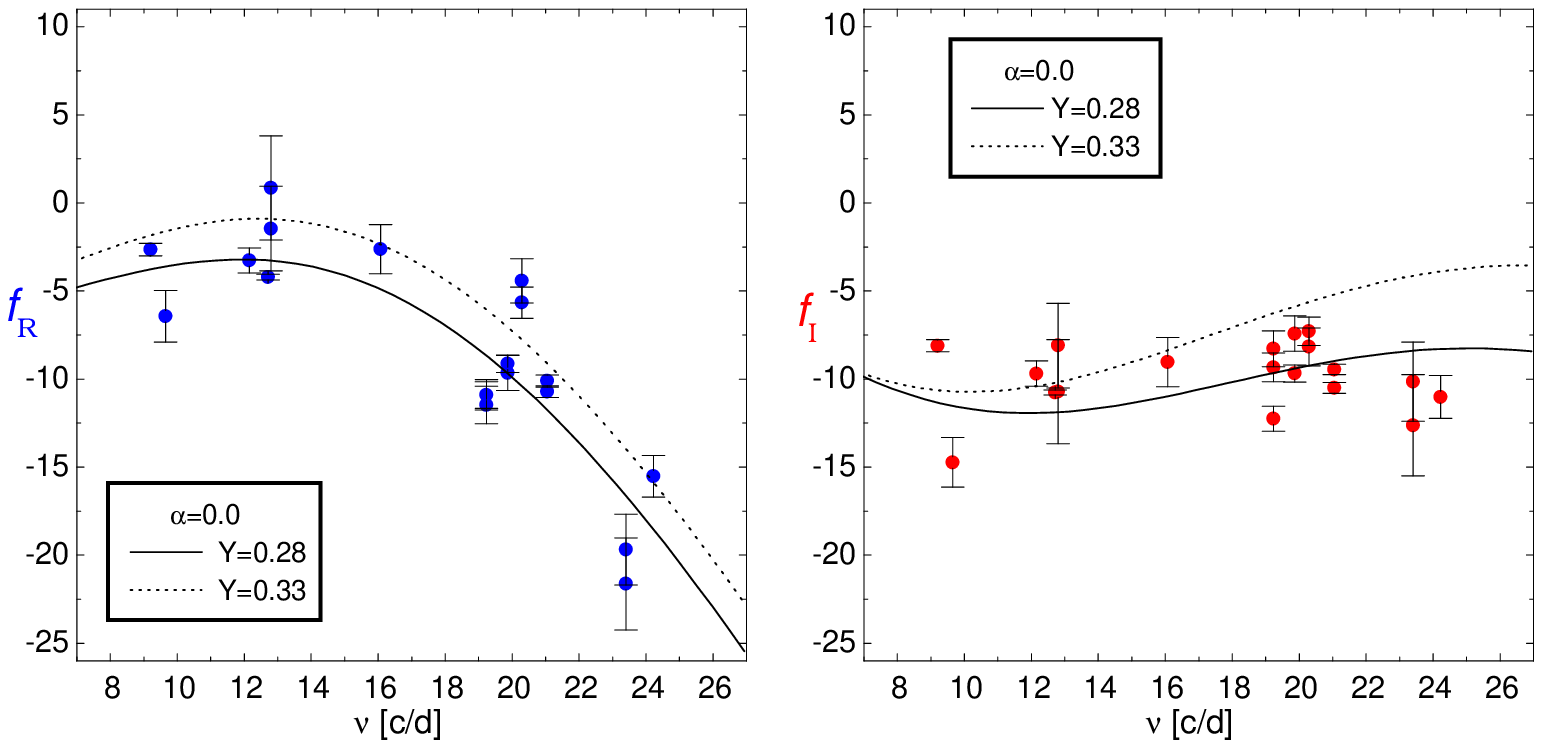}
  \caption{Comparison of the empirical values of $f$
      (dots with error bars) with the theoretical ones calculated for
      $\alpha=0.0$ and two values of the helium abundance: $Y=0.28$ and $Y=0.33$.}
  \label{fig:fparam}
\end{figure*}

\section{Constraints on stellar convection}

It has been shown in Paper I that calculated values of $f$ are
very sensitive to how convection transport is included. Thus, a
comparison with the corresponding empirical values provides a test
of the adopted description of convection.

\subsection{A simplistic approach}

Here, as in Paper I, we begin with the same naive, but
commonly adopted approach i.e. the standard mixing-length theory
(MLT) and the convective flux freezing approximation. FG Vir is a
relatively hot $\delta$ Sct star and, unless the MLT parameter is
unrealistically large, there are two unconnected convective
layers, one associated with H and the other with HeII ionization
zone. The adopted approximation is indeed very bad only in the H
ionization zone, where convective transport may dominate. In
Fig.\,5, we compare empirical values of $f$ with those calculated
upon above approximations. The empirical $f$'s depend only weakly
on adopted values of $T_{\rm eff}$ and $L$. The plotted values
were obtained adopting $M=1.85 M_{\odot}$, $\log T_{\rm
eff}=3.860$ and $\log L=1.170$. These parameters are consistent
with mean photometric data for FG Vir and evolutionary models, and
they lead to $\chi^2 < 1.6$ as obtained with our method for most
of the peaks. The empirical $f$'s are relatively insensitive also
to the choice of $\ell$ which in some cases is ambiguous. For the
plot, we use $\ell$'s corresponding to $\chi^2$ minima. The
calculated $f$'s are even less $\ell$-dependent.

We can see that there is a good agreement between the empirical
and the theoretical values for the models calculated with
$\alpha=0.0$. This is pleasing because our approximations are
irrelevant in the limit of totally inefficient convection.
Calculated $f$'s shown in Fig.\,5 were obtained with the OPAL
opacities (Iglesias \& Rogers 1996) assuming the standard
Population I composition ($Y=0.28, Z=0.02$). We found that results
remain essentially unchanged with the OP opacities (Seaton \&
Badnell 2004, Badnell et al. 2005, Seaton 2005). Also a different
value of metal abundance, $Z$, has only a minimal effect. Only
the change in helium abundance seems to matter a little, as it is
shown in Fig.\,6.

We found that our standard models calculated with $\alpha=0.0$
predict instability of all modes considered in this paper. We take
it as a support for a low value of the MLT parameter and for the
adopted effective temperature. However, in the complete
oscillation spectrum of FG Vir there are peaks extending up to
nearly 45 c/d (Breger et al. 2005). These small amplitude peaks
with frequencies above 25 c/d cannot be explained in terms of
unstable low-degree modes. Above  25 c/d, we found  only
instability of high-degree ($\ell >60$) f-modes. However, to
explain the observed amplitudes in terms of such modes, we would
have to postulate their large intrinsic amplitudes ($\epsilon$)
implying relative temperature fluctuations $>1$, locally in the
hydrogen ionization zone. Therefore, it is unlikely that such
modes may explain the high frequency peaks in FG Vir.  More
likely, these peaks arise from the second order effect of lower
frequency modes leading to peaks at harmonic or combination
frequencies in the oscillation spectrum. The original modes might
not be detected as peaks in the power spectrum if the aspect angle
is close to the node of the spherical harmonic.

\subsection{A more advanced approach}

As an alternative to the simplistic models in Sect.\,4.1, we
considered more advanced models which estimate the turbulent
fluxes by means of a nonlocal time-dependent generalization of the
mixing-length formulation by Gough (1977a, 1977b). In this formulation
there are two more parameters, $a$ and $b$, which control
respectively the spatial coherence of the ensemble of eddies
contributing to the turbulent fluxes of heat, $F_{\rm c}$, and
momentum (also known as turbulent pressure), $p_{\rm t}$, and the
degree to which the turbulent fluxes are coupled to the local
stratification. Roughly speaking, the latter two parameters
control the degree of ``nonlocality'' of convection: low values
imply highly nonlocal solutions, and in the limit
$a,b\rightarrow\infty$ the system of equations formally reduces to
the local formulation (except near the boundaries of the
convection zone, where the local equations are singular). Further
computational details are described by Balmforth~(1992) and Houdek
et al. (1999).

Results presented in Fig.\,7 are for different values of the
mixing-length parameter $\alpha$ but for fixed $a^2=900$ and
$b^2=2000$. From models computed with different values for
$a$ and $b$ we concluded that the $f$ values are rather
insensitive to the choice of $a$ and $b$.
For the remaining convection parameters that are included in
a mixing-length formulation (e.g. anisotropy parameter, see
Gough, 1977b) we assumed values that are consistent with the
formulation by B\"ohm-Vitense (1958).
In the local limit ($a,b\rightarrow\infty$) and for $p_{\rm t}=0$
we obtained for the stellar models approximately the same depth of
the convection zone at constant $\alpha$ between the two formulations
of Sect.\,4.1 and Sect.\,4.2.
The differences in the fractional convective heat flux between the
two convection formulations, depicted in Fig.\,8 for different
values of $\alpha$, are predominantly a result of the effects of
``nonlocality'' and turbulent pressure $p_{\rm t}$.

Let us summarize the differences between the codes used to obtain
the results presented in Sect.\,4.1 (Fig.\,5) and in Sect.\,4.2
(Fig.\,7). In the former case the nonlocal effects of convection
and turbulent pressure were neglected in constructing the
envelope models. There is also a difference in the low-temperature
opacities. In the former case the Alexander \& Ferguson (1994)
and in the latter the Kurucz (1992) opacities were employed.
Moreover, the code that was used to produce the results
in Fig.\,7 assumed the generalized Eddington approximation to
radiative transfer, whereas the diffusion approximation was
assumed in producing the results in Fig.\,5.

In the pulsation calculations that lead to the $f$ values, Gough's
treatment (Fig.\,7) included the perturbations of the convective
heat flux and that of the turbulent pressure, both of which have
been neglected in Sect.\,4.1.

In spite of these differences both cases favour a small
mixing-length parameter ($\alpha\la0.5$), though the models of
Sect.\,4.2 (Fig.\,7), which include convection dynamics, are in
reasonable agreement for a broader range of $\alpha$ values
($0.25\la\alpha\la1.0$). The large differences in $f$ between the
$\alpha=0$ and $\alpha=1$ case depicted in Fig.\,5 must
predominantly result from the convective flux freezing
approximation.

As for mode instability, the results obtained with both convection
formulations are very similar. The advanced approach, which includes
convection dynamics, finds unstable modes for frequencies up to about
25\,c/d (i.e. up to the fifth radial mode) and for models with $\alpha$
values between 0.25 and 1.5. A similar limiting frequency value of about 
25\,c/d is found with the simplistic approach and with $\alpha\simeq0$. 
A significantly larger range of unstable modes is predicted, up to
a frequency of about 31.5\,c/d, for models with $\alpha=1$. Such a large 
value for $\alpha$ is, however, excluded with the simplistic approach 
(see Fig.\,5).

\begin{figure*}
 \includegraphics[width=175mm, height=8cm]{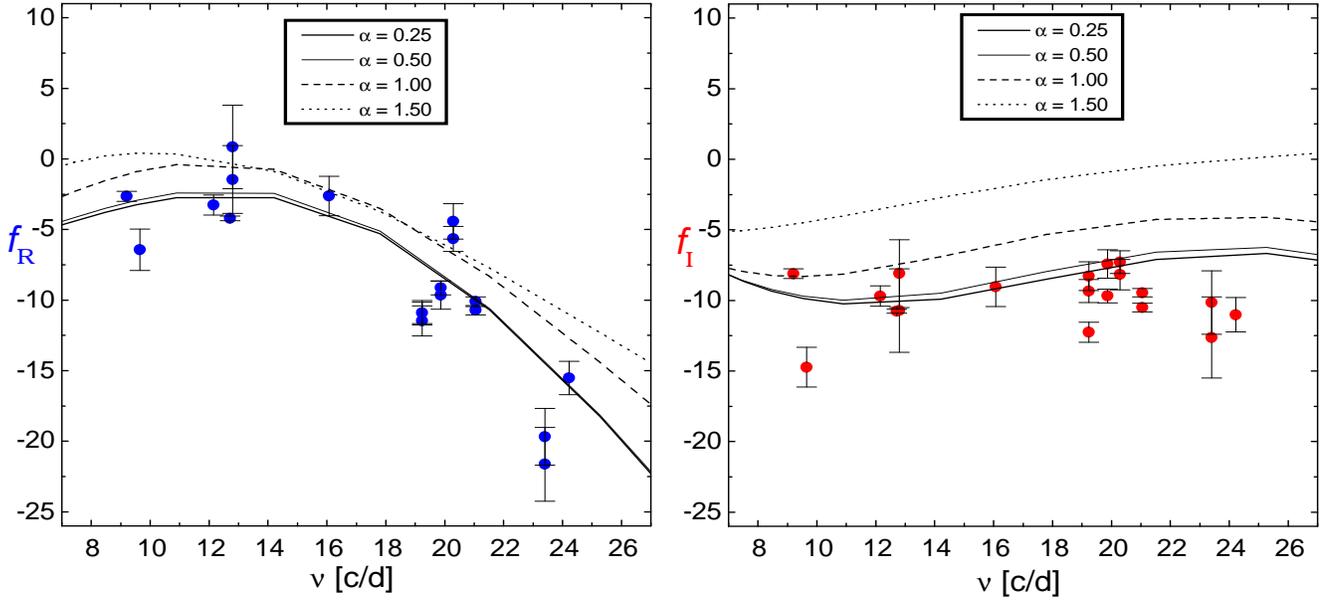}
  \caption{Similar as Fig.\,5 but the theoretical values of $f$ were obtained
           from models using a nonlocal, time-dependent formulation of the
           mixing-length theory (Gough 1977a, 1977b).}
  \label{fig:fparam}
\end{figure*}
\begin{figure}
 \includegraphics[width=87mm, height=9cm]{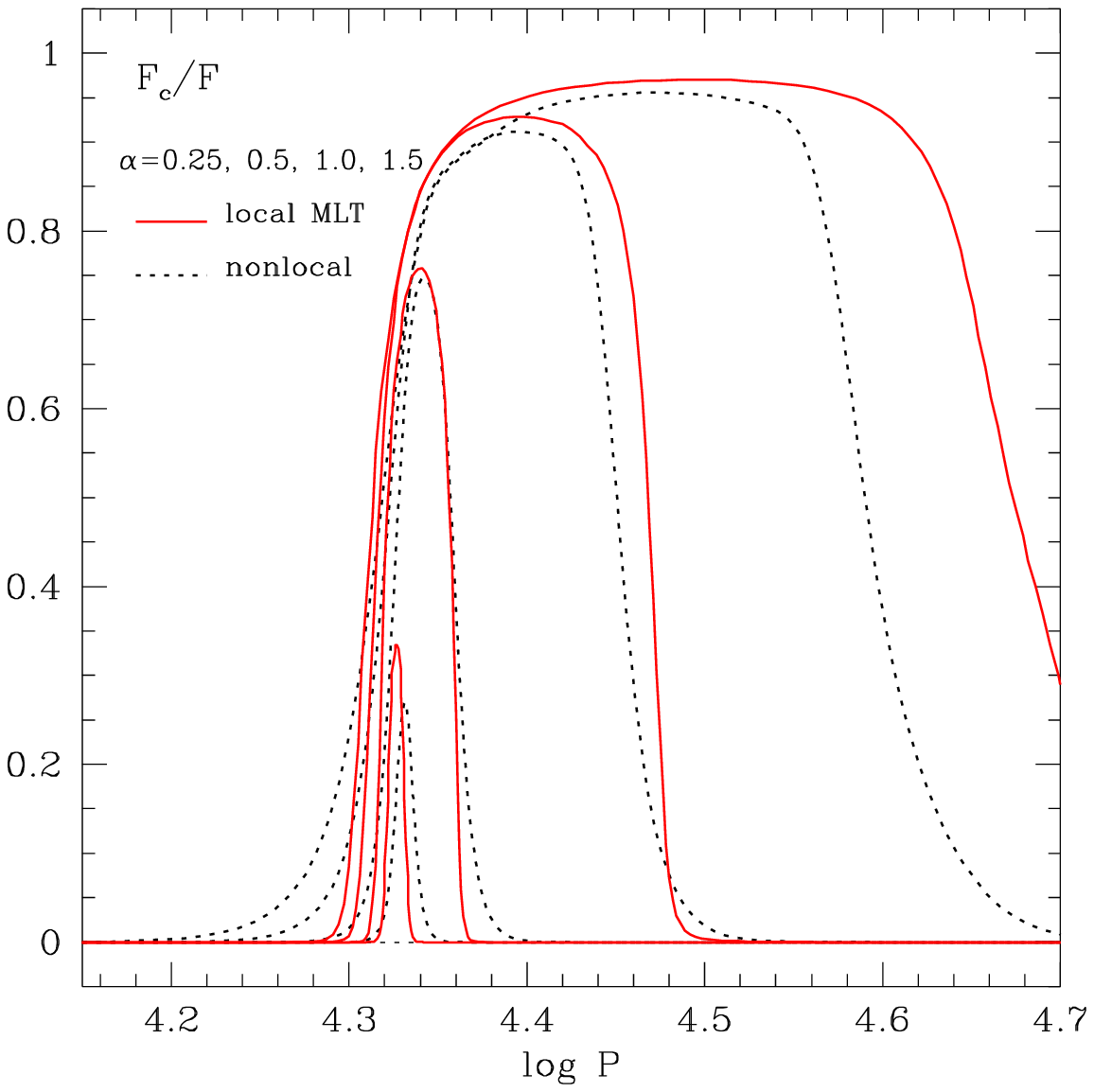}
\caption{Comparison of the fractional heat flux carried by convection
    for various values of the mixing-length parameter, $\alpha$, in the
    standard and in the Gough's nonlocal, time-dependent convection formalisms.
    Results are plotted against total pressure $P$. The maxima of $F_{\rm c}/F$
    are approximately at the centre of the hydrogen ionization zones.} 
  \label{fig:fparam}
\end{figure}

\section{Conclusions}

Using simultaneous photometric and spectroscopic data on twelve
modes excited in FG Vir, we determined their spherical harmonic
degrees, $\ell$, and complex parameters $f$, which link the
surface flux variation to the displacement. In six cases, the
$\ell$ identification is unique at the 80\,\% confidence level. In
all twelve cases, modes with degree  $\ell\ge 3$ are excluded at a
very high confidence level.

The fit of the pulsation data imposes a stringent constraint on
atmospheric parameters, like the effective temperature, $T_{\rm eff}$,
microturbulence velocity, $\xi_{\rm t}$, and metallicity, [m/H].
From the data of the dominant peak in the oscillation spectrum, we
inferred that $T_{\rm eff}$ should be close to the cooler end of
the allowed range defined by the mean colours. For the microturbulent
velocity we found $\xi_{\rm t}\approx 4$ km$\,$s$^{-1}$ if [m/H]$\approx0.0$
was assumed. In the case of low amplitude modes, more accurate
observations and measurements in more passbands are needed
for constraining the atmospheric parameters.

Two of the uniquely identified modes are radial. However, if
effects of rotation are ignored, the observed period ratio is in
conflict with calculated values in standard stellar models
consistent with mean parameters.  We see the best chances for
resolving the discrepancy by taking into account effects of
rotational mode coupling. We will explore this possibility in the
future.

We compared $f$ values inferred from data of twelve pulsation
modes of FG Vir with theoretical values from nondadiabatic
pulsation calculations which assumed various models for convection.
The twelve modes cover a broad range of frequencies. We found good
agreement over the whole frequency range with models for which
convection dynamics was neglected and for which inefficient
convection $(\alpha\simeq0)$ was assumed.
If, however, convection dynamics is included in the model calculations
the results are in reasonable agreement with the data also for larger
values of $\alpha$, though they are still substantially smaller than
for a calibrated solar model.

\begin{acknowledgements}
JDD and GH are grateful to J{\o}rgen Christensen-Dalsgaard 
for instructive discussions during their visit at the Institute 
of Physics and Astronomy in Aarhus.
JDD thanks the Foundation for Polish Science for supporting
her stay in the Copernicus Astronomical Center in Warsaw.
The work was supported by Polish KBN grant  No. 1 P03D 021 28.
GH acknowledges support by the UK Particle Physics and Astronomy
Research Council. The work of MB and WZ has been supported by
the Austrian Fonds zur F\"{o}rderung der wissenschaftlichen Forschung,
grant number P17441-N02.
\end{acknowledgements}

\end{document}